\documentclass[aps,prd,floatfix]{revtex4}
\usepackage{amsmath,amsfonts}
\usepackage[dvips]{graphicx}

\begin{document}

\title{Dark energy FRW cosmology - dynamical system reconstruction}
\author{Marek Szyd{\l}owski}
\affiliation{Astronomical Observatory, Jagiellonian University, Orla 171, 30-244 Krak{\'o}w, Poland}
\affiliation{Complex Systems Research Centre, Jagiellonian University, Reymonta 4, 30-059 Krak{\'o}w, Poland}

\author{Aleksandra Kurek}
\affiliation{Astronomical Observatory, Jagiellonian University, Orla 171, 30-244 Krak{\'o}w, Poland}

\begin{abstract}
We develop a simple method of dark energy reconstruction using a geometrical form of the luminosity-distance relation. In this method the FRW dynamical system with dark energy is reconstructed instead of the equation of state parameter. We give several examples which illustrate the usefulness of our method in fitting the redshift transition from the decelerating to accelerating phase as the value of the Hubble function at the transition.
\end{abstract}

\maketitle

\section{Introduction}
Recent observations of distant SNIa and other astronomical observations of WMAP, SDSS, Chandra X-ray, etc. indicate that current Universe is experiencing an accelerated expansion. There are different cosmological models which explain this phenomenon by postulating the hypothesis that the Universe is filled by fluid which violates the strong energy condition (so called dark energy) or by postulating that the Friedmann equations do not describe well the evolution of the Universe. Due to the lack of a satisfactory cosmological model (while the simplest $\Lambda$CDM model seems to be the best one we have still some problems with it) the model-independent reconstruction of the expansion history of the Universe and the nature of dark energy seems to be important. These reconstruction method can be divided into parametric and non-parametric approach \cite{Sahni:2006}. Here we consider the reconstruction of dynamics in terms of single potential function using the relation between the luminosity distance and the Hubble function. This reconstruction allow us to find the moment of transition from decelerating to accelerating phase of expansion as well as to probe the evolution of the dark energy.
\section{Reconstruction of the potential function of FRW dynamical system}     
Our Universe in large scale structure is homogeneous and isotropic, therefore its geometry in a good approximation can be described by the Robertson-Walker metric. Assuming general relativity we can characterize evolution of the Universe in terms of only single function - scale factor $a(t)$ as a function of cosmic time $t$. The scale factor obeys the Raychaudhuri equation (also called acceleration equation)
\begin{equation}\label{eq:1}
\frac{\ddot{a}}{a}=-\frac{1}{6}(\rho+3p),
\end{equation}
where dot denotes differentiation with respect the cosmological time $t$ and $\rho=\rho(t)$, $p=p(t)$ are energy density and pressure of the perfect fluid - source of gravity. 

We assume that both $\rho$ and $p$ depends on $t$ through the scale factor, i.e. $\rho=\rho(a(t))$ and $p=p(a(t))$. Then we can simply find first integral of Raychaudhuri equation, called as Friedmann equation 
\begin{equation}
H^2(a) \equiv \frac{\dot{a}^2}{a^2} = \frac{\rho}{3}-\frac{k}{a^2},
\end{equation}
where $H$ is Hubble function, energy density $\rho$ satisfies conservation condition $\dot{\rho}=-3H(\rho+p)$, $k=0, \pm 1$ is curvature constant.

Equation (\ref{eq:1}) can be rewritten to the form of a Newtonian dynamical system
\begin{equation}\label{eq:3}
\ddot{a}=-\frac{\partial V(a)}{\partial a}, \ \ \ \frac{\dot{a}^2}{2} + V(a)\equiv 0,
\end{equation}
where $V=V(a)$ is the potential function which characterizes a model and dot means here differentiation with respect rescalled time variable $\tau$: $t \rightarrow \tau$: $|H_{0}| dt = d\tau$.

In this interpretation the Friedmann first integral plays the role of energy integral. For example for the concordance $\Lambda$CDM model potential function assumes the following form 
\begin{equation}
V(a)=-\frac{1}{2} \sum_{i={m,k,\Lambda}} \Omega_{i,0}a^{-1-3w_{i}},
\end{equation}
where $\Omega_{i,0}$ are density parameters for each fluid $\Omega_i=\frac{\rho_i}{3H_0^2}$ defined at the present epoch labelled by $0$ index; $w_m=0$ for dust matter, $w_k=-\frac{1}{3}$ for curvature fluid and $w=-1$ for the cosmological constant; $p_i=w_i(a)\rho_i$ in general and $a_0=1$ is chosen for the value of the scale factor at present.

Let us note that in a general case if equation of state is parameterized by the scale factor $a$ then all dynamics is characterized in terms of a single potential function of $a$ (or redshift $z$: $1+z=a^{-1}$). Of course this function has the general form
\begin{equation}
V(a)=-\frac{1}{2}\Omega_{\textrm{eff}}(a)a^2,
\end{equation} 
where $\Omega_{\textrm{eff}}$ is effective energy density parameter for total fluids.\\
Since $V=V(a)=-\frac{(H/H_0)^2a^2}{2}$ the potential function $V(z)$ can be immediately reconstructed from SNIa data. Such a reconstruction is possible due to a simple relation between $H(z)$ and luminosity distance for the flat Universe 
\begin{equation}\label{eq:6}
c \left [ \frac{d}{dz} \left ( \frac{d_L(z)}{1+z} \right ) ^{-1} \right ] =H(z).
\end{equation}
Due to the fact that we know $d_L(z)$ for the discrete values of $z$ a kind of smoothing procedure is needed to obtain function $V(z)$. This can be done by approximating the luminosity distance function by some fitting formula. Here we consider different ansatz for $\frac{H_0}{c}\frac{d_L(z)}{1+z}\equiv f(\bar{\theta},z)$ function 
\begin{enumerate}
 \item polynomial models $f(\bar{\theta},z)=\sum_{i=0}^k a_iz^i$, where $k={1,\dots,4}$,
 \item Pade-type ansatz $f(\bar{\theta},z)=P_{n,m}$, where  $(n,m)=\{(1,1),(1,2),(1,3),(2,1),(2,2),(3,1)\}$,
as well as $f(\bar{\theta},z)=2\left[\frac{x-a_1\sqrt{x}-1+a_1}{a_2+a_3\sqrt{x}+2-a_1-a_2-a_3}\right]$, where $x=z+1$ which was proposed in \cite{Saini2000},
 \item model analysed in \cite{Chiba:2000}: $f(\bar{\theta},z)=\eta(1)-\eta(y)$, where $\eta(y)=2\alpha\left[y^{-8}+\beta y^{-6} + \gamma y^{-4} +\delta y^{-2} + \sigma\right]^{-\frac{1}{8}}$ with $y=(1+z)^{-\frac{1}{2}}$.
\end{enumerate}
Additionally we consider following constraints 1. $d_{L}(z=0)=0$, 2. $H(z=0)=H_0$.\\

Finally we have chosen this model which is the best one in the light of SNIa data \cite{Davis:2007, Riess:2007, Wood:2007} in the Bayesian framework meaning, i.e. with the greatest value of the posterior probability \cite{Jeffreys:1961, Szydlowski:2006xx}. Here we used the BIC quantity \cite{Schwarz:1978} as an approximation to the evidence and assumed equal values of prior probabilities for all models.

\begin{figure}[ht]
\centering
\includegraphics[height=.26 \textheight]{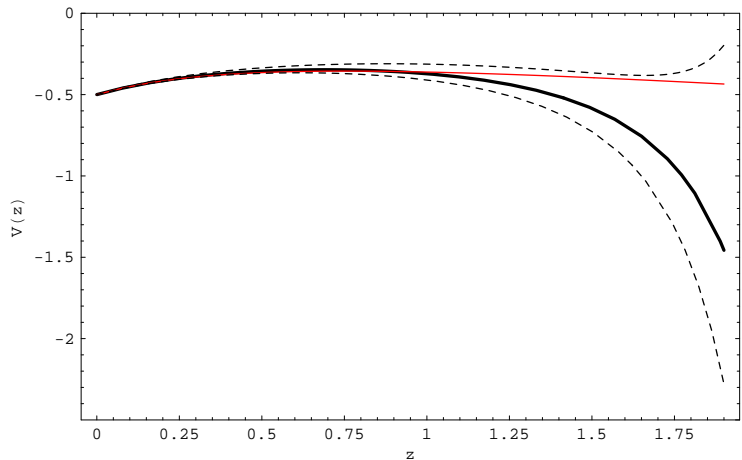}
\caption{$V(z)$ for $f(\bar{\theta},z)=a_2z^2+z$ (solid, black line) together with 1$\sigma$ confidence level (dashed lines) with $a_2=-0.21_{-0.015}^{+0.023}$ and for $\Lambda$CDM model (red line) with $\Omega_{m,0}=0.27$.}
\label{fig:1}
\end{figure}

After such an analysis we conclude that the best parametrization (model) is $f(\bar{\theta},z)=a_2z^2+z$ and we have used this one in the later analysis. Result of reconstruction the potential function is presented in Figure \ref{fig:1}.\\

The moment of transition ($a_{T}$, $z_{T}$) is related with the maximum of the potential function so it can be easily found. Results for the reconstructed potential are as follows $a_{T}=0.59$ ($z_{T}=0.69$), $V(z_T)=-0.35$ and $\frac{H(z_T)}{H_0}=1.98$. These outcomes can be compared with the values obtained for the potential of the best fit $\Lambda$CDM model $a_{T}=0.57$ ($z_{T}=0.75$), $V(z_T)=-0.35$ and $\frac{H(z_T)}{H_0}=1.47$. \\
   
In the next section we consider the scalar field $\phi$ with the potential $U(\phi)$ and show a simple method of obtaining $U(\phi(a))$ directly from $V(a)$.
   
\section{Potential of a scalar field from the potential of the FRW dynamical system}

Let us consider a single scalar field with potential $U(\phi)$ in the FRW model. They are described by the natural lagrangian function
\begin{equation}
\mathcal{L}_{\phi}=\frac{1}{2} g_{\mu \nu}\partial^{\mu}\phi \partial^{\nu} \phi -U(\phi).
\end{equation} 
We assume that a scalar field is homogeneous and minimally coupled to gravity, so kinetic energy term is of the form $T=\frac{1}{2}\dot{\phi}^2$, $\dot{}\equiv \frac{d}{dt}$. \\
The equation of motion for the field satisfies the Klein-Gordon equation \begin{equation}
\ddot{\phi}+3H\dot{\phi} +\frac{\partial U}{\partial \phi}=0.
\end{equation}
The previous equation is equivalent to a perfect fluid with energy density and pressure as follows
\begin{equation}\label{eq:10}
\rho_{\phi}=\frac{1}{2}\dot{\phi}^2+U(\phi);\ \  p_{\phi}=\frac{1}{2}\dot{\phi}^2-U(\phi).
\end{equation}
If we assume that both scalar field as well as potential depends on time through the scale factor then coefficient of state $w_{\phi}=w_{\phi}(a)$ can be reconstructed directly from (\ref{eq:10}), namely
\begin{equation}
w_{\phi}(a)=\frac{T+U}{T-U}.
\end{equation}
And vice-versa if we parametrize $w_{\phi}(z)$ by redshift then both kinetic term and potential can be reconstructed from simple relation
\begin{eqnarray}
\frac{1}{2}\left( \frac{d\phi}{dt} \right)^2=\frac{1}{2}\left(1+w_{\phi}(a)\right)\rho_{\phi}(a),\label{eq:12}\\
U(\phi(a))=\frac{1}{2}\left(1-w_{\phi}(a)\right)\rho_{\phi}(a).\label{eq:13} 
\end{eqnarray}
The evolution of quintessence scalar field is governed by adiabatic conservation condition 
\begin{equation}
\dot{\rho}_{\phi} +3H(\rho_{\phi}+p_{\phi})=0,
\end{equation}
which solution is 
\begin{equation}
\rho_{\phi}=\rho_{\phi,0}a^{-3}\exp \left [ -3 \int_{1}^{a}\frac{w_{\phi}(a)}{a}da \right]= 
\rho_{\phi,0}a^{-3(1+\bar{w}_{\phi}(a))},
\end{equation}
where $\bar{w}_{\phi}(a)=\frac{\int w_{\phi}(a)d(\ln a)}{\int d(\ln a)}$. \\

Let us consider flat FRW cosmological model filled by dust matter and dark energy in the form of quintessence scalar field. Therefore we have 
\begin{equation}
\rho_{eff}=\rho_{m,0}a^{-3}+\rho_{\phi,0}E(a),
\end{equation}
where $E(a)=a^{-3} \exp \left [-3 \int_{1}^{a}\frac{w_{\phi}(a)}{a}da \right ]$.

Becausse we assume that matter and dark energy sectors does not interact the potential function is additive, i.e.
\begin{equation}
V(a)=V_m(a)+V_{\phi}(a),
\end{equation}
where $V_m(a)=\Omega_{m,0}a^{-1}$; $\Omega_{m,0}=\frac{\rho_{m,0}}{3H_0^2}$; $V_{\phi}(a)=\Omega_{\phi,0}a^2E(a)$; $\Omega_{\phi,0}=\frac{\rho_{\phi,0}}{3H_0^2}$. It is a simple consequence of the fact that
\begin{eqnarray} V_{\phi}=-\frac{\rho_{\phi}a^2}{6}=-\frac{1}{6}a^2\left(\frac{1}{2}\left(\frac{d\phi}{dt}\right)^2+U(\phi)\right)= \nonumber \\
=\frac{1}{6}a^2\left(\frac{1}{2}(1+w_{\phi})\rho_{\phi}+\frac{1}{2}(1-w_{\phi})\rho_{\phi} \right )= \nonumber \\
=-\frac{1}{2}\Omega_{\phi}a^2.
\end{eqnarray}
From the assumption of $k=0$ we have the Friedmann first integral in the form
\begin{equation}
\rho_m+\rho_{\phi}=\rho_{\textrm{eff}}=3H^2
\end{equation}
and therefore relation (\ref{eq:12}) determine dependence of $\phi(a)$:
\begin{equation}\label{eq:20}
\left( \frac{d\phi}{da}\right)^2= \frac{\rho_{\phi}(a)(1+w_{\phi}(a))}{(aH(a))^2}.
\end{equation}

\begin{figure}[ht]
\includegraphics[height=.2\textheight]{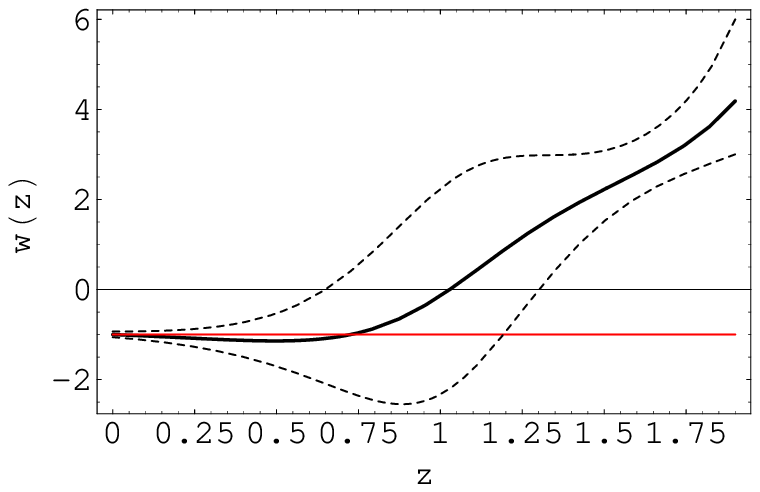}
\includegraphics[height=.2\textheight]{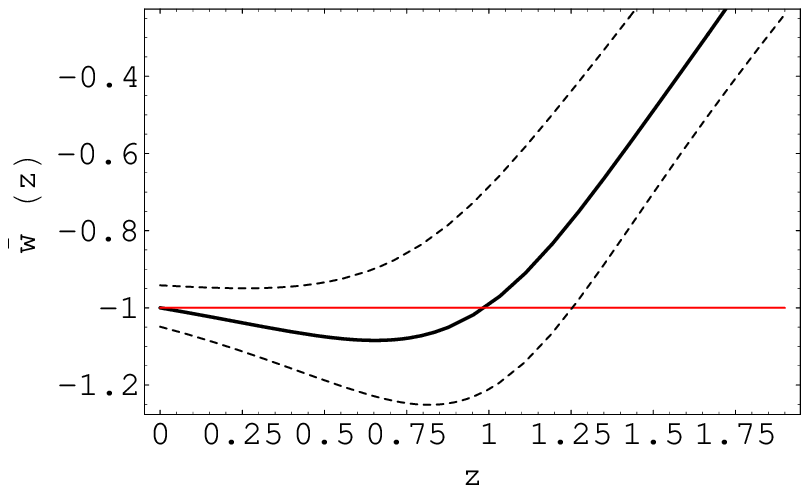}
\caption{$w_{\phi}(z)$ and $\bar{w}_{\phi}(z)$ for $f(\bar{\theta},z)=a_2z^2+z$ (solid, black lines) together with 1$\sigma$ confidence level (dashed lines) with $a_2=-0.21_{-0.015}^{+0.023}$, $\Omega_{m,0}=0.27$ and $w(z)$ for the $\Lambda$CDM model (red lines).}
\label{fig:2}
\end{figure}

\begin{figure}[ht]
\includegraphics[height=.2\textheight]{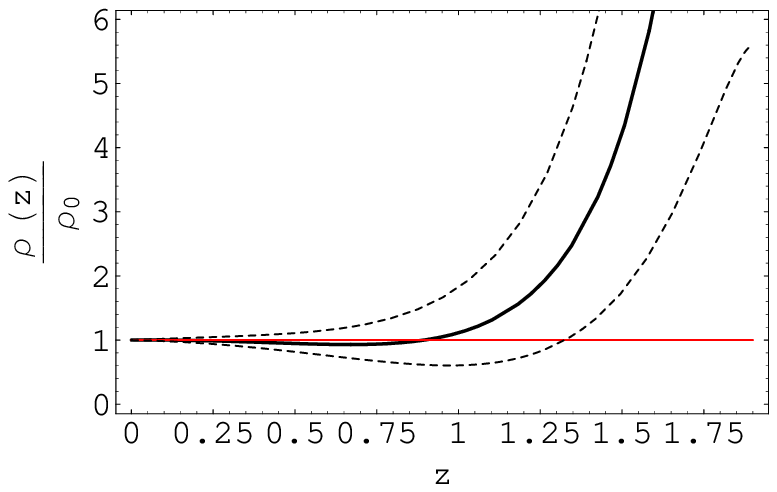}
\includegraphics[height=.2\textheight]{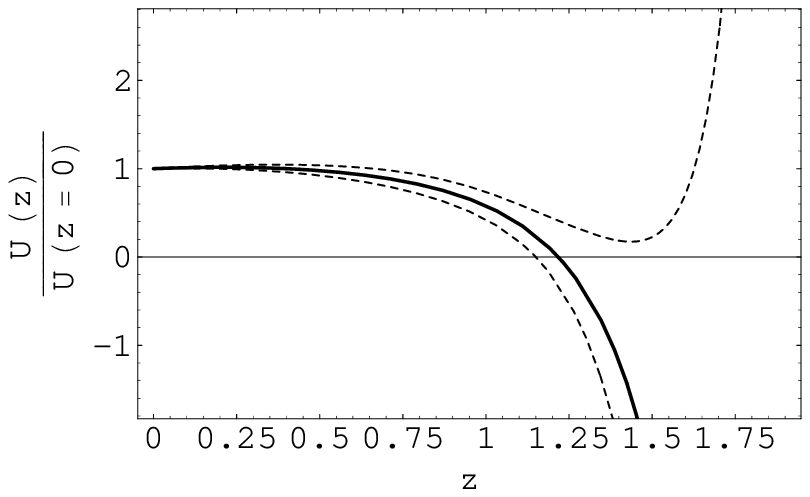}
\caption{$\frac{\rho_{\phi}(z)}{\rho_{\phi,0}}$ and $\frac{U(z)}{U(z=0)}$ for $f(\bar{\theta},z)=a_2z^2+z$ (solid, black lines) together with 1$\sigma$ confidence level (dashed lines) with $a_2=-0.21_{-0.015}^{+0.023}$, $\Omega_{m,0}=0.27$ and $\frac{\rho_{\Lambda}(z)}{\rho_{\Lambda,0}}$ for $\Lambda$CDM model (red line). }
\label{fig:4}
\end{figure}

One can check that (\ref{eq:1}) implies that the dark energy coefficient equation of state, $w_{\phi}(a)$, is related to the FRW potential $V$ as follows
\begin{equation}\label{eq:21}
w_{\phi}(a)=-\frac{1}{3}\left[ 1-\frac{d(\ln|V_{\phi}|)}{d(\ln a)}\right]=-\frac{1}{3}\left[ 1-\frac{d(\ln|(V-V_m)|)}{d(\ln a)}\right],
\end{equation}
where $V=-\frac{1}{2}\Omega_{eff}(a)a^2$, $V_m=-\frac{1}{2}\Omega_{m,0}a^{-1}$ (see Figure \ref{fig:2}).
Hence after substitution (\ref{eq:21}) into (\ref{eq:20}) we obtain relation $\phi=\phi(a)$.\\
Let us note that $\rho_{\phi}$ appeared in (\ref{eq:20}) can be integrated in an exact form in term of the potential $V$, namely $\rho_{\phi}=\rho_{\phi,0}a^{-3}a^{-3\bar{w}(a)}$, where 
\begin{equation}\label{eq:22}
-3\bar{w}_{\phi}(a)=1+\frac{\ln [|V(a)+\frac{1}{2}\Omega_{m,0}a^{-1}|]}{\ln a} - \frac{\ln [ |V(a=1)+\frac{1}{2}\Omega_{m,0}|)}{\ln a},  
\end{equation}
(see Figure \ref{fig:2}) which leads to 
\begin{equation}\label{eq:23}
\rho_{\phi}(a)=\rho_{\phi,0}\left( \frac{V(a)+\frac{1}{2}\Omega_{m,0}a^{-1}}{V(a=1)+\frac{1}{2}\Omega_{m,0}} \right) a^{-2},
\end{equation}
(see Figure \ref{fig:4}).
The analogous procedure of the reconstruction one can perform for the potential of the scalar field $U(\phi)$. For this aim we substitute $w_{\phi}(a)$ from formula (\ref{eq:21}) and $\rho_{\phi}(a)$ from (\ref{eq:23}) into (\ref{eq:13}). Finally we obtain the following formula for reconstruction potential of scalar field as a function of scale factor:
\begin{equation}\label{eq:24}
U(a)=U(a=1)\frac{a^{-2}\left[ 4V(a)+\frac{3}{2}\Omega_{m,0}a^{-1}+a\frac{d V(a)}{d a}\right ]}{ 4V(a=1)+\frac{3}{2}\Omega_{m,0}+\frac{d V(a)}{d a}|_{a=1}},
\end{equation}
(see Figure \ref{fig:4}).
Both formulas can be rewritten to the form containing redshift $z$ instead the scale factor.  

\section{Conclusion}
Different cosmological models can be observationally differentiate in terms of the potential function of the dynamical system. We reconstructed the potential function of the FRW dynamical system directly from SNIa observations using the relation between the luminosity distance and the Hubble function and the parametric method of smoothing the function $d_L(z)$. As one concludes the shape of the reconstructed potential is similar to the shape of the $\Lambda$CDM model potential for small redshift ($z<1$). We compute the moment of transition using the reconstructed potential as well as the potential of the $\Lambda$CDM model. We also consider the cosmological model with a single, homogeneous and minimally coupled to gravity scalar field $\phi$ with the potential $U(\phi)$. We derive the relation between $w_{\phi}(a)$/$\rho_{\phi}(a)$/$U(a)$ function and the potential function of the FRW dynamical system $V(a)$ which allow us to reconstruct the evolution of such functions without making assumptions on the form of the $w_{\phi}$ and $\rho_{\phi}$. We compare evolution of such quantities with the evolution predicted by the $\Lambda$CDM model.      

\begin{acknowledgments}
This work has been supported by the Marie Curie Actions
Transfer of Knowledge project COCOS (contract MTKD-CT-2004-517186).
\end{acknowledgments}


\end{document}